# The Emory Optic Nerve Head Atlas – Using 3D Anatomical Mapping to Study Optic Neuropathies with an Initial Focus on Glaucoma

Thanadet Chuangsuwanich,[1,2] Patipol Tiyajamorn,[1,2] Yibo Chen,[1,2] Michael Dattilo,[1] Deepta Ghate,[1] Nancy Newman,[1,3] Valérie Biousse,[1,3] and Michaël J.A. Girard[1,2,4,5]

[1]Department of Ophthalmology, Emory University School of Medicine, Atlanta, Georgia USA
[2]Ophthalmic Engineering & Innovation Laboratory, Emory University School of Medicine, Atlanta, Georgia USA.
[3]Department of Neurology, Emory University School of Medicine, Atlanta, Georgia USA
[4]Department of Biomedical Engineering, Georgia Institute of Technology/Emory University, Atlanta, GA, USA
[5]Emory Empathetic AI for Health Institute, Emory University, Atlanta, GA, USA

**Short Title:** The Emory Optic Nerve Head Atlas

| | |
|---|---|
| **Word count:** | 4,752 (manuscript text) |
| | 493 (abstract) |
| **Tables:** | 2 |
| **Figures:** | 6 |
| **Supplementary Material:** | 1 |

**Corresponding Authors:**

Michaël J.A. Girard, Ophthalmic Engineering & Innovation Laboratory, Emory Eye Center, Emory School of Medicine, Emory Clinic Building B, 1365B Clifton Road, NE, Atlanta GA 30322.
Email: mgirard@ophthalmic.engineering

## Abstract

**Purpose**: To develop the first 3D optic nerve head (ONH) atlas using AI-based registration and evaluate its use in: (1) atlas-adjusted retinal nerve fiber layer (RNFL) analysis for glaucoma diagnosis, and (2) strain-based assessment of glaucoma severity.

**Methods**: Large-scale OCT datasets were registered using REFLECTIVITY-generated tissue segmentations. A healthy atlas (n=460) and glaucoma atlases for mild (n=852), moderate (n=640), and severe (n=546) disease were built using an AI-driven algorithm incorporating structural and biomechanical constraints. Atlas quality was evaluated using inter-layer contrast and agreement between warped scans and the template. Atlas-adjusted RNFL thickness was extracted by mapping fixed 1.5 mm Bruch's membrane opening (BMO) based reference points from the atlas to each subject and compared with standard subject-specific RNFL measurements. Effective strain relative to each atlas was computed by morphing subjects to the atlas template. Strain and thickness features were used to classify glaucoma severity.

**Results**: The healthy atlas captured normative ONH anatomy, while glaucoma atlases showed stage-dependent changes (cup deepening, prelaminar thinning, rim narrowing). Registration accuracy was high (NCC=0.86±0.05; Dice=0.90±0.02), with strong inter-layer contrast (0.40±0.03). Atlas-adjusted RNFL thickness improved separation between healthy and mild glaucoma (Cohen's d=0.77 vs. 0.72). Classification using atlas-adjusted profiles outperformed native measurements (AUC=0.771 vs. 0.757). Mean effective strain differed significantly across severity groups (ANOVA p<0.05), and CNNs trained on strain maps achieved strong classification (AUC=0.79).

**Conclusion**: The ONH atlas provides a population-derived structural reference enabling anatomically consistent thickness mapping and strain-based characterization. It enhances diagnostic discrimination, supports severity classification, and offers a framework to study glaucomatous biomechanics.

# Introduction

In many areas of medicine, complex anatomy is studied using a population atlas – a three-dimensional reference that represents the average structure of a healthy organ or tissue.[1-4] Brain imaging provides the best-known example: Individual MRI scans are typically aligned and 'morphed' to such a standard reference (e.g. Montreal Neurological Institute [MNI] atlas),[5] allowing comparisons of brain structure and function to be made at matching anatomical regions across individuals. Such atlases reduce the impact of natural anatomical variability and have become essential for identifying early disease-related changes, integrating multimodal data, and developing more reliable biomarkers.[6,7] Despite the central role of the optic nerve head (ONH) in glaucoma – the leading cause of irreversible blindness worldwide[8] – and in other optic neuropathies, no comparable atlas has yet been developed for the eye.

The absence of an ocular atlas is particularly relevant because ONH anatomy varies widely among individuals, even in healthy eyes. Disc size, tilt, and cup depth can differ substantially from one person to another.[9] Yet, key glaucoma biomarkers – such as retinal nerve fiber layer (RNFL) thickness, a primary indicator of axonal loss measured with optical coherence tomography (OCT) – are typically sampled at fixed locations across all eyes.[10] In standard practice, RNFL thickness is measured along a 3.4 mm diameter circle centered on the disc, regardless of whether the eye or disc is small or large.[10] As a result, measurements taken at a fixed radius do not necessarily correspond to the same anatomical regions across individuals. This inconsistency means that normal anatomical variability can mask subtle, early glaucomatous changes and reduce diagnostic precision. A population-based ONH atlas could help overcome this limitation by standardizing structural comparisons across eyes.

Beyond standardizing structural measurements, an atlas can also be used to quantify how each individual eye deviates from the average healthy anatomy. In essence, each ONH can be aligned (or "morphed") to match a healthy reference atlas. The deformation required for this alignment can be expressed mathematically as "strain", a concept borrowed from biomechanics that describes how much a shape must stretch, compress, or shear to fit another.[11] In this context, strain does not represent a physical load on the tissue but rather the geometric difference between an individual ONH and the healthy template. Mapping this strain across the ONH provides an intuitive visualization of localized structural deviations and allows subtle disease-related remodeling to be quantified. Thus, atlas-based strain offers a second application of the framework, complementing RNFL standardization by providing a direct, spatially resolved measure of structural abnormality in glaucoma.

In this manuscript, we aimed to develop the first 3D atlas of the ONH using advances in AI-based registration techniques. We evaluated its utility in two glaucoma-related domains: **(1)** improving glaucoma diagnosis through atlas-adjusted analysis of RNFL thickness; and **(2)** assessing glaucoma severity using atlas-derived strain mapping to identify localized structural deviations linked to disease progression.

# Methods

## *Overview of Atlas Construction*

The goal of this study was to construct a 3D atlas of the ONH that represents the population-average anatomy of both healthy and glaucomatous eyes using an image registration framework (Figure 1). At first glance, the simplest way to build such an atlas would be to directly average all OCT scans voxel-by-voxel (for example, taking the mean or median intensity at each location). However, even when scans are aligned using a common landmark such as the BMO, this naïve averaging approach produces a blurry and anatomically indistinct volume. Differences in eye size, tilt, tissue curvature, and local morphology accumulate across subjects, causing anatomical structures to "wash out" rather than reinforce one another (Figure 2).

To address this, we used an AI–based Iterative Deformable Image Registration (IDIR) method designed specifically to bring all scans into the same spatial reference. The concept is simple: instead of averaging misaligned volumes, we gradually deform each scan so that its anatomy matches a shared reference. We begin with an initial template created by taking the mean of all unregistered scans. Although this starting template is blurry, it is an unbiased representation of the population and serves as the baseline that all scans will move toward. In each iteration, IDIR aligns every scan to the current template, warps them into this shared space, and re-averages the results to produce an updated, sharper template. Repeating this process steadily improves alignment and ultimately yields a stable "converged" template that best reflects the population average. This final template is the 3D ONH atlas. A detailed flow chart of this process is given in **Appendix A**.

To support this alignment, each OCT scan was also segmented to identify major tissue layers and anatomical boundaries (Figure 1). These segmentations were converted into 3D point clouds, which provide clear geometric landmarks that complement the OCT intensity and help guide the registration toward anatomically meaningful matches across individuals.

By combining OCT intensity, tissue segmentation, and biomechanical mapping within this iterative IDIR framework, the resulting atlas provides a detailed and consistently aligned representation of ONH structure across subjects. The complete pipeline is described in the following sections.

### *Data Acquisition*

We accessed a large OCT imaging database at Emory University through CONTINUUM, a secure, HIPAA-compliant cloud platform for clinical imaging research (IRB-approved). Subject identifiers were used only for medical record review to confirm diagnostic and demographic information; all subsequent image analyses were performed on deidentified OCT volumes.

Glaucoma subjects were identified using the ICD-10 diagnostic code H40.1 (open-angle glaucoma), and diagnoses were confirmed by expert neuro-ophthalmologists through chart review. Exclusion criteria included other optic neuropathies (e.g., compressive or ischemic optic neuropathy, optic neuritis), retinal diseases (e.g., diabetic retinopathy, macular degeneration), or any prior intraocular surgery that could alter optic nerve head anatomy. OCT volumes were acquired using the CIRRUS HD-OCT system (Carl Zeiss Meditec), with each scan centered on the optic nerve head (ONH) and covering a 6 mm × 6 mm field of view.

Volumes were sampled at an in-plane resolution of 30 $\mu$m and an axial resolution of 2 $\mu$m, yielding voxel dimensions of 200 × 200 × 1024 (transverse × transverse × axial). A total of 2,038 glaucomatous eyes were included and stratified by severity based on standard visual field criteria: 852 mild (mean deviation [MD] > –6 dB), 640 moderate (–6 ≥ MD > –12 dB), and 546 severe (MD ≤ –12 dB).

For the healthy cohort, 460 eyes were selected from individuals undergoing routine ophthalmic evaluations with no history or clinical evidence of optic neuropathy. Healthy eyes were defined as those with normal optic disc appearance, intact neuroretinal rim, and no signs of glaucomatous or other optic nerve pathology. Inclusion criteria required normal intraocular pressure (≤ 21 mmHg) and best-corrected visual acuity of 20/25 or better. Exclusion criteria included any ocular disease (e.g., glaucoma, optic neuritis, ischemic or compressive optic neuropathy), refractive error exceeding ± 6 diopters, prior intraocular surgery, or systemic conditions known to affect the optic nerve (e.g., diabetes with retinopathy, multiple sclerosis). All subjects were confirmed to be glaucoma-free based on comprehensive clinical review and evaluation by expert neuro-ophthalmologists. Race was self-reported in the medical record.

### AI-based Segmentation of ONH Tissues Enabling 3D Point Cloud Representation of ONH Structure

We used REFLECTIVITY (Abyss Processing Pte Ltd, Singapore) to automatically segment all baseline OCT volume scans of the ONH and to align the volumes to the BMO plane. To ensure consistency, all eyes were standardized to the right-eye orientation for atlas generation. The following ONH tissue groups were segmented (Figure 1a): the RNFL and prelaminar tissue, the ganglion cell–inner plexiform layer, all other retinal layers, the retinal pigment epithelium with Bruch's membrane and BMO points, choroid, the OCT-visible portion of the LC, and the OCT-visible portion of the peripapillary sclera. Each ONH was then represented as a 3D point cloud derived from the segmentations, following our previously established approach.[12,13]

The resulting segmentations, boundary point clouds, and BMO landmarks were used as input for our deep learning–based atlas generation framework.

### Deep Learning–Based Image Registration for Atlas Generation

To create a population-based atlas, each individual OCT volume must be precisely aligned to a common reference. The process of geometrically transforming one image to match another by minimizing intensity differences is known as image registration.

Traditional 3D registration methods such as digital volume correlation (DVC) have been successfully implemented with OCT data of the ONH.[14-18] However, DVC requires extensive preprocessing[11] and is sensitive to image noise and motion artifacts, which are common in clinical OCT scans - particularly from older systems such as Cirrus OCT (as used herein). It also assumes small, continuous deformations between samples, a condition rarely met during atlas construction.

To overcome these limitations, we implemented an AI-based registration approach called Implicit Deformable Image Registration (IDIR).[19] IDIR is a recent deep learning framework that learns how to deform one 3D OCT volume to match another by directly modeling the geometric transformation between them. Instead of requiring large datasets for training, IDIR operates in an unsupervised manner, using a simple neural network that predicts a continuous deformation field for each image pair.

In essence, IDIR "morphs" each subject's ONH to align with the reference template while preserving anatomical realism. The model is constrained to produce biomechanically plausible deformations - that is, smooth shape changes without tissue folding or distortion. To further improve accuracy, we incorporated tissue boundary information from the segmentations (represented as 3D point clouds) to help guide the alignment process.[20]

This AI-driven approach enables accurate, tissue-level registration across thousands of OCT volumes while avoiding the computational and noise-related limitations of conventional methods such as DVC. Detailed model architecture, optimization settings, and loss functions are provided in **Appendix B**.

## Initial Template Generation

To initialize ONH atlas generation, we first constructed a reference template for each cohort (healthy, mild, moderate, & severe glaucoma). A pixel-wise mean intensity image was computed from all OCT volumes in a cohort after a rigid alignment using the BMO center as a reference. In parallel, we generated a segmentation template by taking the voxel-wise mode across all tissue segmentations, thereby preserving the most frequently occurring label at each voxel and providing a noise-resistant anatomical prior. These averaged templates served as the initial references for subsequent registration and atlas construction. An example of the initial templates (both image and segmentation) for severe glaucoma subjects is shown in **Figure 2**.

## Iterative Atlas Construction

We constructed the ONH atlases for each cohort (healthy, mild glaucoma, moderate glaucoma and severe glaucoma) using an iterative registration process based on the IDIR framework. The procedure began with an initial reference template, to which each subject's OCT scan was aligned. At each iteration, IDIR registered every subject's scan to the current template, generating deformation fields that were used to warp both the OCT volumes and their corresponding segmentations into the template. An example of a morphed image with its associated landmarks is shown in Figure 1c.

The newly aligned scans were then averaged to create an updated template, and the process was repeated. With each cycle, the template became increasingly representative of the population as registration accuracy improved and inter-subject variability decreased.

Convergence was defined as the point at which further iterations produced negligible improvement in alignment. This was assessed by monitoring image similarity (normalized cross-correlation, NCC) and segmentation overlap (Dice coefficient) between successive template versions. NCC ranges from $-1$ to $1$, with values closer to $1$ indicating stronger similarity between image intensity patterns, whereas Dice scores range from $0$ to $1$, with values closer to $1$ reflecting greater overlap of corresponding tissues. Typically, NCC values $>0.9$ and Dice scores $>0.85$ are considered excellent for image registration. The process was considered converged when changes in both metrics were less than $0.1$ across two consecutive iterations.

At convergence, the resulting template represents the final **atlas** - an anatomically consistent, population-average models that captures the shared structural features of the ONH in each cohort.

### *Quantifying 3D Deformation Metrics from the IDIR Field*

To characterize how a given ONH morphs into another, we quantified structural differences using deformation metrics. In biomechanics, these metrics describe how tissues stretch, compress, or shear under physical load. Here, we repurposed them to capture anatomical variation between eyes – reflecting geometric deviations from a healthy reference atlas rather than deformations induced by IOP.

The **deformation gradient** provides a local description of how small regions within the ONH are stretched, compressed, or rotated during the transformation from an individual ONH to the atlas. From this, the **Jacobian determinant** can be derived, which quantifies local volume changes: values greater than one indicate expansion, while values less than one indicate compression. The **strain tensor** summarizes the overall magnitude and direction of this local deformation, and from it we compute **the effective strain**, a single scalar value representing the total extent of shape change regardless of direction. We have applied effective strain metric extensively in our prior work to characterize regional ONH biomechanics.[11,12,21-23] The definition of these terms are given in **Appendix C**.

### *Assessment of Atlas Structural Fidelity*

To evaluate registration performance, we computed the average NCC and Dice between each healthy subject's registered volumes/segmentations and the final healthy atlas. Qualitative visual inspection was first performed by ONH anatomy experts (T.C., M.G.). To confirm that common structural features were accurately captured, we further assessed the structural fidelity of the atlas by quantifying the inter-layer contrast – a measure the absolute difference in mean OCT reflectivity between adjacent tissue layers. Higher inter-layer contrasts indicate clearer delineation of tissue boundaries, while lower values reflect blurring or reduced layer separation. Higher values of inter-layer contrast (typically > 0.4 or 0.5) indicating sharper, well-defined anatomical separation compared to lower values which suggest blurring. For this, a 20 × 40 pixel region of interest (ROI) was sampled in the central slice of the volume, approximately 1.5 mm from the BMO

center, and inter-layer contrast was estimated using the equation described in our previous work.[24]

We also characterized atlases ONH morphometrics using REFLECTIVITY. From the atlases segmentation, we extracted structural parameters including prelaminar depth, prelaminar thickness, minimum rim width (MRW), RNFL thickness, and BMO area.

### ***Comparison between Atlas-Adjusted RNFL Thickness Profile vs Conventional RNFL Thickness Profile***

To demonstrate the utility of the atlas in standardizing ONH geometry, we compared RNFL thickness measurements obtained from two approaches. In the conventional fixed-diameter approach, tissue thickness was measured directly in each subject at a fixed 3 mm diameter from their individual BMO center.[25] In the atlas-adjusted approach (Figure 3a), reference points were first defined at a fixed 1.5 mm radius from the BMO in the healthy atlas. These points were then mapped to each subject's ONH using the deformation fields generated by IDIR, and tissue thickness was extracted at the corresponding anatomical locations.

Group comparisons across glaucoma severity stages (healthy, mild, moderate, severe) were performed to assess whether the atlas-adjusted RNFL thickness profile (by quadrant) improved clinically relevant differences between the groups. Pairwise two-sample t-tests were conducted (healthy–mild, mild–moderate, moderate–severe), and effect sizes were quantified using Cohen's d. This analysis allowed us to compare the atlas-adjusted method with the conventional fixed-diameter approach and assess the potential utility for anatomically standardized, cross-subject measurements.

### *Comparing Atlas-Adjusted RNFL Thickness Profile and Conventional RNFL Thickness Profile in Classifying Early Glaucoma from Healthy*

We conducted a proof-of-concept classification experiment to distinguish healthy eyes from those with early glaucoma using a simple multilayer perceptron (MLP) comprising two hidden layers with 128 and 64 units, respectively. The input feature was the RNFL thickness sampled at 1° intervals around the optic nerve head. Model performance was evaluated using five-fold cross-validation, and the area under the receiver operating characteristic curve (AUC) was reported for models trained on both native and atlas-adjusted RNFL thickness profiles. This experiment provided an initial demonstration of whether atlas-adjusted RNFL thickness profiles could improve disease classification accuracy in a nonlinear deep learning framework.

### *Using Individual-to-Altas Strain as a Structural Deviation Metric*

Leveraging our atlases, we explored the use of strain fields to compare deformations across disease-specific atlases and individual patient scans (**Figure 4a**).

For each subject, strains were computed relative to the corresponding disease-stage atlases (healthy, mild glaucoma, moderate glaucoma, and severe glaucoma). Effective strain, a scalar measure of tissue deformation, highlights localized regions of swelling, compression, or shear and provides intuitive, spatially resolved visualization through color mapping.

In addition, we quantified individual-to-healthy atlas strain, which represents the structural deviation of each subject from the normative healthy atlas. This approach enables characterization of complex, localized 3D structural changes across disease states. For example, Figure 4a illustrates a patient scan morphed to the healthy atlas, revealing subject-specific regions of localized strain. Group-level comparisons were performed by computing the mean and standard deviation of effective strain maps within each cohort, followed by pairwise t-tests to assess differences between groups.

We further evaluated the potential of effective strain as an input for automated classification. A 3D CNN was trained to perform four-class classification (healthy, mild, moderate, and severe glaucoma) using effective strain volumes. The network architecture consisted of three convolutional layers (64, 128, 256 filters; kernel size 3; padding 1) with batch normalization and max-pooling at each stage. To manage computational cost, we sampled 300 subjects per group, with 80/20 train/validation split and 5-fold cross validation, and model performance was assessed by mean AUC across folds.

## Results

### *Subjects' demographics*

A total of 2,498 subjects were included in this study, consisting of 460 healthy controls and 2,038 patients with glaucoma. The healthy control group had a mean age of 51 ± 17 years and 68% of the cohort were female. The glaucomatous cohort was stratified into 852 mild, 640 moderate, and 546 severe cases based on visual field mean deviation (MD) values at the time of the OCT scan. The mean age of glaucoma subjects was 65 ±

15 years, and 54% of the cohort were female. MD values ranged from –1.0 dB (mild) to –36.2 dB (severe), with a cohort-wide average of 9.0 ±8.2 dB.

The cohort was composed of approximately 55% African American, 41% White, and 4% from other ethnic backgrounds (including Hispanic, Asian, and multiracial). This distribution reflects the populations in the southeastern United States, particularly in the Atlanta area.

A summary of demographic and clinical characteristics, stratified by glaucoma severity, is presented in Table 1.

### *IDIR Achieves Anatomically Accurate and Computationally Efficient Registration*

The IDIR framework produced anatomically accurate, smooth, and physically plausible deformations (Figure 1c). It is also computationally efficient, requiring approximately three minutes for each registration. Registration accuracy was quantified using normalized cross-correlation (NCC) to assess intensity similarity and Dice similarity coefficients to measure tissue overlap. Across all cohorts, the mean NCC between warped subject volumes and the final atlas was 0.86 ± 0.05, and the mean Dice score was 0.90 ± 0.02 across all segmented layers, indicating excellent performance in registration accuracy.

Visual inspection further confirmed precise alignment, including consistent registration of the Bruch's membrane opening, lamina cribrosa contours, and prelaminar shape. Representative examples of patient scans morphed to the atlas are shown in the appendix, illustrating how IDIR preserves fine anatomical detail while accommodating biological variability.

***Atlases Capture Structural Fidelity and Biological Variability Across Glaucoma Severity***

Mild, moderate, and severe glaucoma atlases were well representative of their respective disease stages, as confirmed by quantitative morphometric analysis using the REFLECTIVITY software. The minimum rim width (MRW), a clinically recognized indicator of neuroretinal rim thinning, decreased from 340 µm in healthy eyes to 226, 126, and 113 µm in mild, moderate, and severe glaucoma, respectively. The prelaminar tissue became progressively thinner and more posteriorly displaced, while the BMO area increased, indicating overall disc enlargement. Together, these trends demonstrate that the constructed atlases maintain high structural fidelity while accurately reflecting biologically meaningful variability associated with glaucoma progression. A summary of these morphometric characteristics is provided in Table 2.

The constructed ONH atlases exhibited high anatomical fidelity, showing distinct tissue boundaries and consistent alignment across all severity levels (Figure 5). Quantitative analysis of inter-layer contrast, a proxy for structural fidelity, showed the highest values in healthy eyes (0.54 ± 0.20 units), followed by mild (0.45 ± 0.22), moderate (0.44 ± 0.22), and severe glaucoma (0.37 ± 0.24). For comparison, a standard high-quality OCT scan typically exhibits a contrast of approximately 0.4.[26] The progressive decline in contrast with disease severity reflects increasing anatomical distortion and reduced signal separation across neural tissue layers.

### *Atlas-Adjusted RNFL Thickness Profiles Enhance Discrimination Between Healthy and Early Glaucoma*

Atlas-adjusted RNFL thickness profiles provided clearer separation between healthy and mild glaucoma groups compared to the conventional fixed-diameter profiles. The average adjusted RNFL thickness was 113 ± 25 µm (healthy) vs 93 ± 25 µm (glaucoma), while native measurements were 112 ± 27 µm (healthy) vs. 94 ± 28 µm (glaucoma). Although mean values were similar, adjusted profiles exhibited reduced variability, particularly in the glaucoma group, leading to improved group separation. Specifically, Cohen's d effect size increased from 0.72 to 0.77 with adjustment. Statistical significance also improved, with the t-test p-value shifting from $7 \times 10^{-19}$ (native) to $8 \times 10^{-21}$ (adjusted). Consistent improvements were also observed when comparing mild vs. moderate and moderate vs. severe glaucoma (data not shown).

### *Atlas-Adjusted RNFL Thickness Profiles Enhance Disease Classification*

Models trained on atlas-adjusted thickness profiles achieved higher performance (AUC: 0.77 ± 0.02) compared to those trained on the conventional profiles (AUC: 0.75 ± 0.03) (**Figure 3c**).

### *Strain-Based Atlas Mapping Reveals Stage-Dependent Structural Deviations in Glaucoma*

Average effective strain relative to the healthy atlas increased progressively with glaucoma severity, reflecting stage-dependent deformation from normal anatomy. The mean effective strain was 21 ± 11% for healthy eyes, 25 ± 17% for mild glaucoma, 27 ±

17% for moderate glaucoma, and 30 ± 25% for severe glaucoma (Figure 4b). One-way ANOVA confirmed significant group differences ($p < 0.05$).

To further evaluate the utility of atlas-based measures for downstream AI applications, we employed effective strain (relative to the healthy atlas) as input for classification. Strain maps enabled reliable separation across four glaucoma severity groups. Multiclass models achieved a mean AUC of $0.79 \pm 0.03$, with the clearest discrimination observed between mild and severe glaucoma. These results highlight the promise of strain as a robust, spatially resolved biomarker that complements conventional thickness-based metrics.

We also visualized effective strain distributions across subject groups when morphed to each corresponding atlas (Figure 6). In both healthy and severe cohorts, strain increased progressively as subjects deviated from their own atlas (e.g., healthy eyes showed the highest strain when mapped to glaucoma atlases, and vice versa). Consistently, localized regions of elevated strain were identified at the neuroretinal rim, lamina cribrosa insertion, and prelaminar tissue.

## Discussion

In this study, we developed the first three-dimensional atlases of the ONH from OCT images, leveraging a deep learning-based registration framework to achieve anatomically consistent alignment across subjects. By integrating both intensity- and structure-based constraints with biomechanics-based regularization terms, the framework produced a high-fidelity atlases that preserved anatomical fidelity.  We demonstrated the utility of this framework through two proof-of-concept applications: (1) atlas-adjusted RNFL thickness profiles, which improved separation between glaucoma severity groups and enhanced early disease classification, and (2) individual-to-atlas strain mapping, which revealed stage-dependent strain patterns. Collectively, these results establish a robust foundation for atlas-guided structural phenotyping of the ONH.

Standardization of ocular anatomy remains a major challenge in glaucoma imaging.[27] Conventional metrics such as RNFL thickness are referenced to the center of BMO,[28,29] but variability in disc size, shape, and orientation introduces noise that can obscure disease-related differences.[30] Our results show that atlas-based morphing reduces this anatomical variability, yielding tighter distributions of RNFL thickness in glaucoma and enhancing statistical group separation. Notably, even small improvements in effect size and classification performance can be clinically meaningful, especially in early disease detection where sensitivity is most needed. This approach draws direct parallels to neuroscience, where population-level brain atlases such as the MNI atlas[4] have transformed how we study neuroanatomy and neurodegenerative disease.[6,31] These frameworks enable reproducible mapping of function, pathology, and connectivity across individuals.[4,5] In a similar fashion, the ONH atlas has a potential to become a common reference for ocular structures, supporting consistent cross-subject comparisons, standardization of multi-cohort datasets, and more robust integration of imaging data with functional measures.

Another key innovation of this framework is the ability to compute strain as a measure of deviation from normative anatomy. In our prior works, strain has been used to quantify biomechanical responses of the ONH to changes in intraocular pressure and gaze.[11,21,32] These analyses highlighted how tissues deform between two physiological states and revealed biomechanically vulnerable regions linked to glaucoma. In contrast, the strain measure presented here does not represent a deformation within an individual due to loads (e.g. IOP changes), but instead represents a structural deviation map, quantifying how an individual ONH differs from a population-derived healthy atlas. It is important to note that atlas-based strain does not capture true remodeling, as it does not represent longitudinal changes within the same individual. Instead, it quantifies the overall structural deviation from a normative reference, providing a measure of how an individual's anatomy differs from the healthy standard—analogous to how normative RNFL thickness is used to assess a patient's glaucoma status. In our analysis, effective strain maps revealed severity-dependent increases in deviation from healthy anatomy and consistently highlighted localized regions of structural alteration at the neuroretinal rim, lamina cribrosa insertion, and prelaminar tissue. These spatial patterns align with regions known to be susceptible to glaucomatous damage and corroborate both biomechanical models and histological evidence of ONH remodeling.[33-37] Thus, atlas-based strain could represent a complementary metric for characterizing cumulative structural changes in glaucoma.

We also conducted a preliminary proof-of-concept experiment to test the feasibility of integrating atlas-based measures into an AI-based classification framework. The model was simple - designed not for optimal performance but to evaluate whether these new atlas-derived metrics could enhance prediction in principle. Our results showed that both atlas-adjusted RNFL thickness profiles improved classification performance relative to conventional structural metrics. While these gains were modest, they demonstrate that atlas-based features carry additional discriminative value that can be exploited by a basic neural network. Importantly, we were also able to classify glaucoma severity using atlas-based strain volumes alone. These strain volumes encapsulate rich and spatially interpretable information, making them well suited for more advanced deep learning frameworks, an avenue we plan to explore in future work.

Beyond glaucoma, atlas-based standardization and strain mapping hold promise for a wide spectrum of optic neuropathies, including compressive, ischemic, inflammatory, and hereditary forms. These conditions often exhibit overlapping patterns of structural loss,[38] yet differ in the spatial distribution and magnitude of tissue remodeling. For example, due to the similar nature of optic disc atrophy with cupping, some optic neuropathies such as hereditary or compressive forms might be mistakenly identified as glaucoma.[38] A population-derived 3D atlas could thus provide a unifying framework to disentangle disease-specific deformation signatures. Moreover, as the atlases expand to encompass multiple diseases and severities, they could serve as normative templates for developing cross-disease biomarkers and standardizing structural measurements across imaging centers.

Several limitations warrant consideration. First, inter-individual variability in vessel architecture poses challenges for establishing a unique one-to-one anatomical correspondence of vessel features during atlas registration. Thus, the resulting deformation fields may not represent a strictly anatomically unique mapping, potentially limiting the applicability of our approach to diseases with prominent vessel-specific features. Second, visualization of the LC was limited in many scans, constraining the ability to fully characterize deep ONH structure; future work with swept-source and ultra-widefield OCT could overcome this limitation. Third, while atlas construction and strain computation were computationally efficient relative to traditional methods, real-time clinical translation will require further optimization or cloud-based deployment. Fourth, our study was cross-sectional; no longitudinal data were available to test the sensitivity of atlas-based strain for detecting progression. Demonstrating longitudinal utility will be essential to establish strain as a clinically actionable biomarker. Because racial background was determined from self-reported data, possible variations in optic disc and lamina cribrosa anatomy across racial groups could influence the observed structural patterns. While the diversity of the study population is a strength, this heterogeneity should be considered when interpreting atlas-derived morphological differences. Finally, although we presented proof-of-concept AI applications, broader integration of atlas-based strain with multimodal imaging and longitudinal datasets will be necessary to fully assess its role in progression monitoring and personalized treatment planning.

In summary, we present the first 3D ONH atlas constructed from OCT imaging, establishing a standardized coordinate system for structural comparison and biomechanical characterization. Drawing parallels to brain atlases in neuroscience, which have reshaped the study of neurodegeneration, the ONH atlas and strain framework have the potential to transform our understanding of optic neuropathies and pave the way for precision diagnostics in ophthalmology.

# Acknowledgments

We acknowledge support from **(1)** the Emory Eye Center [Start-up funds, MJAG], **(2)** a grant from the National Eye Institute of the National Institutes of Health (NEI/NIH) – 1R01EY037299-01 [MJAG]; **(3)** the NIH grant P30EY06360 to the Atlanta Vision Community [MJAG, VB, NJN]; and **(4)** a Challenge Grant from Research to Prevent Blindness, Inc. to the Department of Ophthalmology at Emory University [MJAG, VB, NJN, TC].

| Characteristic | Healthy | Mild Glaucoma | Moderate Glaucoma | Severe Glaucoma |
|---|---|---|---|---|
| Age (year) | 51± 17 | 63 ± 15 | 66 ± 15 | 67 ± 14 |
| Sex, female (%) | 68% | 55% | 55% | 50% |
| Visual field, MD (dB) | - | -2.3 ± 6.6 | -8.6 ± 1.73 | -20.4± 5.9 |
| Pattern standard deviation (dB) | - | 3.3 ± 10.1 | 8.1 ± 13.1 | 10.0 ± 8.1 |

**Table 1.** Subjects' demographics.

| Morphological Features | Healthy Atlas | Mild Glaucoma Atlas | Moderate Glaucoma Atlas | Severe Glaucoma Atlas |
|---|---|---|---|---|
| Minimum Rim Width ($\mu$m) | 340 | 226 | 126 | 113 |
| Prelamina depth ($\mu$m) | 13 | 245 | 270 | 400 |
| Prelamina thickness ($\mu$m) | 421 | 202 | 166 | 171 |
| BMO area ($mm^2$) | 1.89 | 1.99 | 1.96 | 2.17 |

**Table 2.** Morphological features of the constructed atlases. Reported values represent measurements obtained from the final atlas of each cohort and are intended to describe the atlas anatomy rather than population-level statistics.

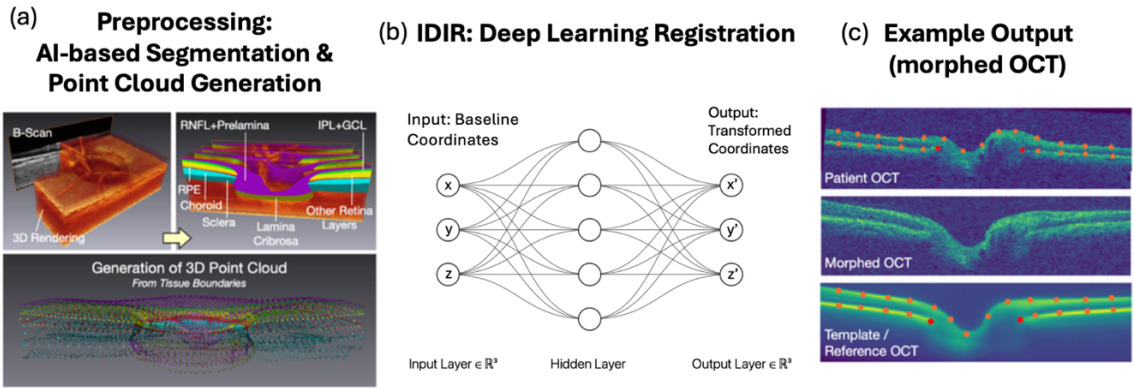

**Figure 1.** Workflow of atlas-based registration using IDIR. **(a)** OCT volumes are segmented using AI methods to delineate tissue boundaries, from which a 3D point cloud representation of the optic nerve head (ONH) is generated. **(b)** The IDIR network takes baseline 3D coordinates (x, y, z) as input and predicts transformed coordinates (x′, y′, z′), enabling nonlinear alignment to a reference template. **(c)** Example output illustrates how a patient OCT is morphed to match the template OCT, with tissue boundaries represented as orange point clouds and Bruch's Membrane Opening (BMO) landmarks shown as red points. These serve as point cloud and landmark constraints, respectively, to ensure anatomically accurate registration.

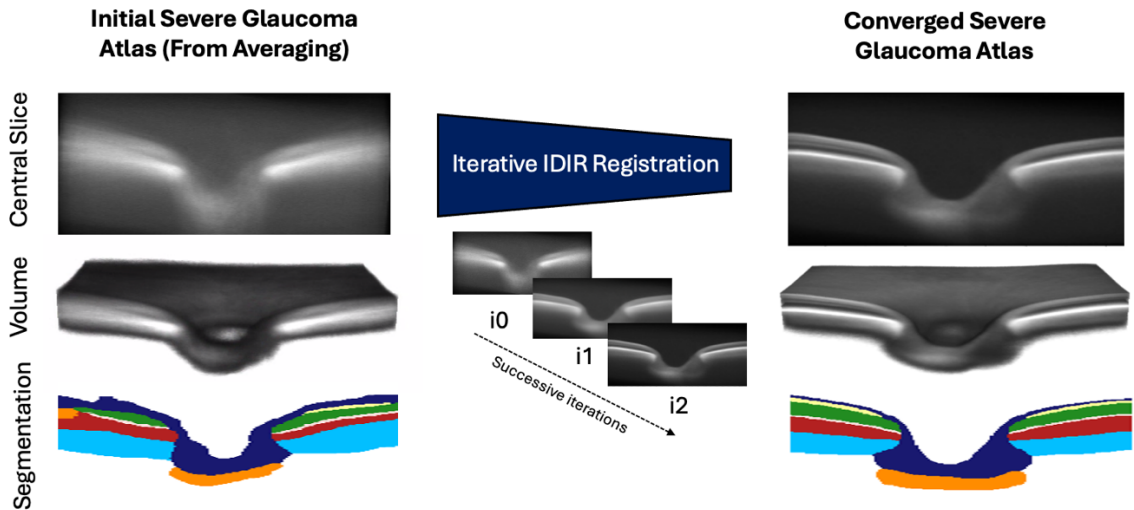

**Figure 2.** Comparison of the initial and converged severe glaucoma atlases. The initial atlas, generated by direct averaging of subject volumes (left), appears blurred with poorly defined tissue boundaries. After iterative deep learning registration, the converged atlas (right) shows sharper anatomical definition in the central slice, 3D volume, and tissue segmentation, producing a more anatomically representative reference of severe glaucoma.

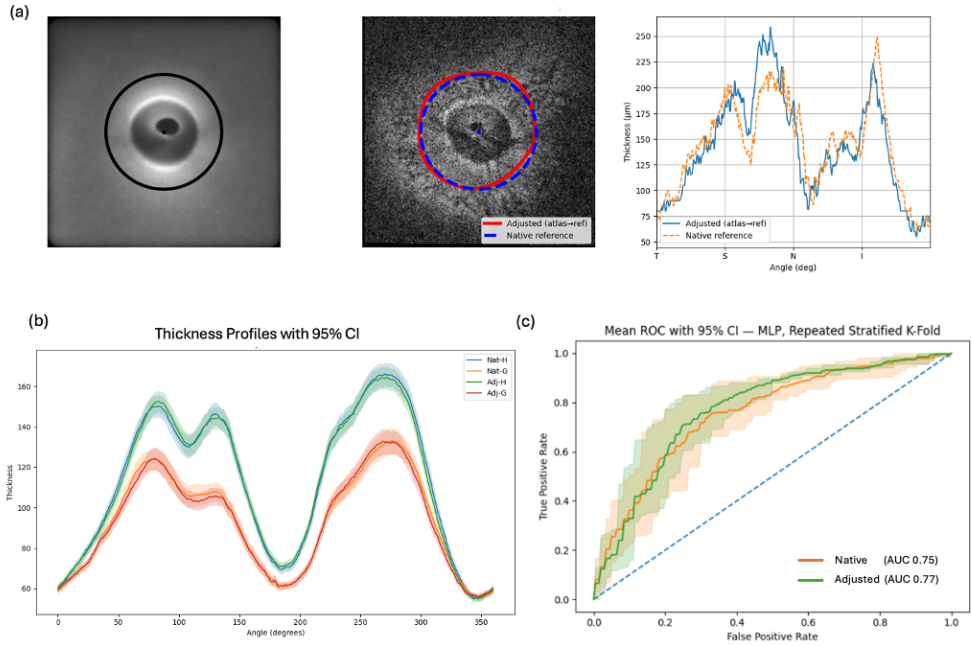

**Figure 3.** Atlas-based adjustment of RNFL sampling relative to the Bruch's Membrane Opening center. The atlas framework was used to standardize the RNFL sampling circle to 1.5 mm centered on the BMO. (a) Example subjects shown with the adjusted sampling relative to the healthy atlas reference and corresponding RNFL thickness profiles. (b) Mean RNFL thickness profiles for all healthy and mild glaucoma subjects. (c) Classification performance (healthy vs. mild glaucoma) using native versus adjusted sampling, demonstrating improved accuracy with atlas adjustment (AUC: 0.757 to 0.771).

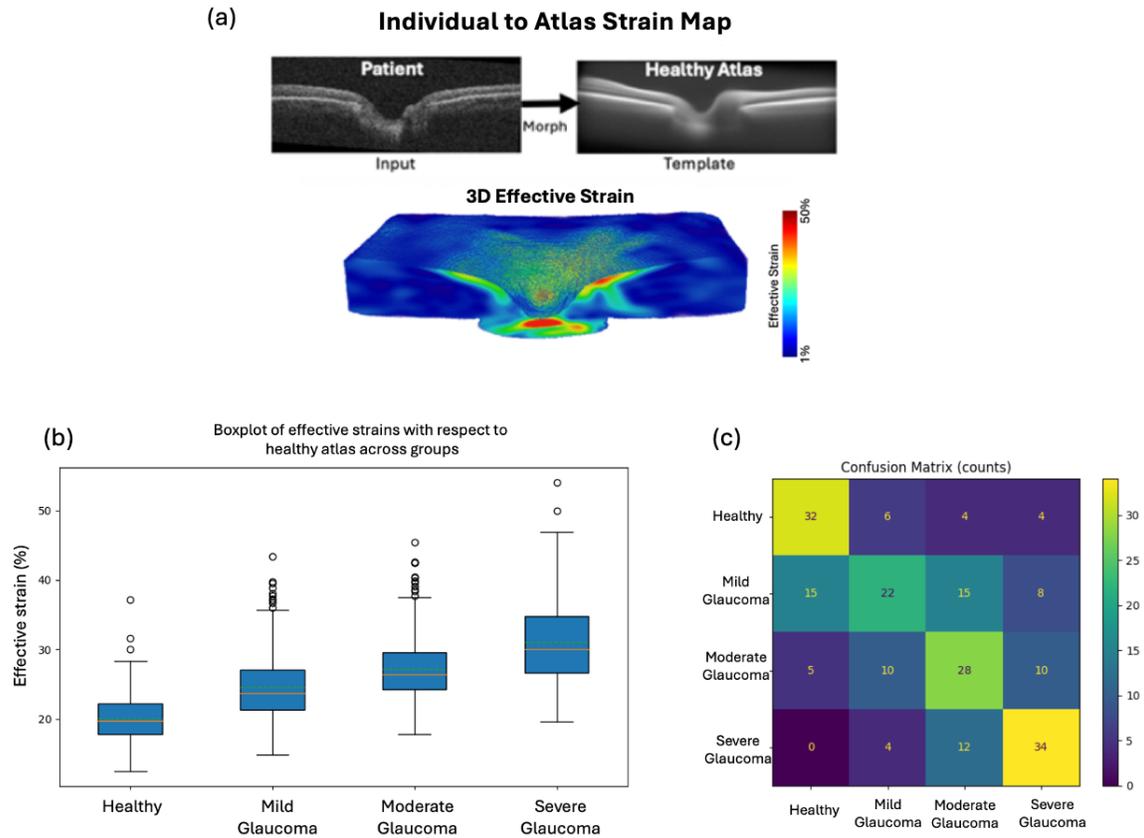

**Figure 4.** Individual-to-atlas strain mapping and group-level analysis. (a) Example of atlas registration, where a patient OCT volume is morphed to the healthy atlas, enabling computation of a 3D effective strain map. (b) Boxplots of effective strain (%) relative to the healthy atlas across groups (healthy, mild, moderate, and severe glaucoma) show a progressive increase in strain, with significant differences observed between groups. (c) Multiclass classification performance using a multilayer perceptron (MLP) with effective strain as input, shown as a confusion matrix, achieved an overall AUC of 0.79.

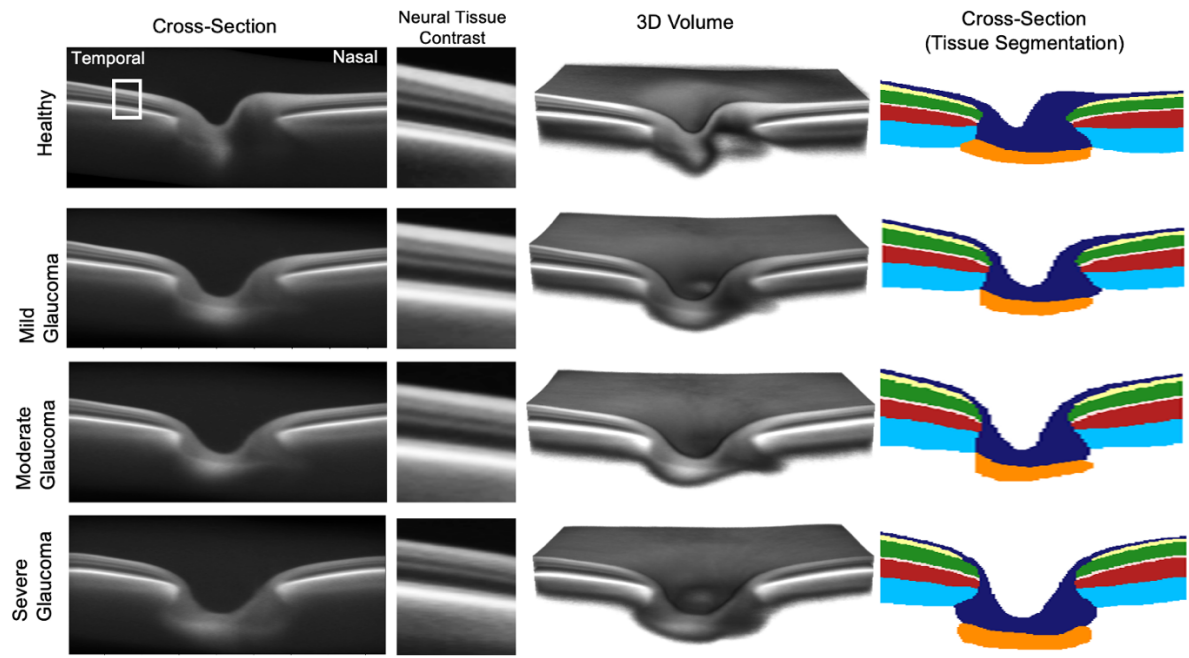

**Figure 5.** Structural features of the ONH atlases across glaucoma severity. Representative OCT-derived cross-sections (left), zoomed neural tissue contrast (second column), 3D reconstructions (third column), and corresponding tissue segmentations (right) are shown for a healthy eye and eyes with mild, moderate, and severe glaucoma. Progressive disease severity is associated with thinning of neural tissues, deepening of the ONH cup, and loss of neuroretinal rim tissue, as captured in both raw cross-sections and segmented tissue layers.

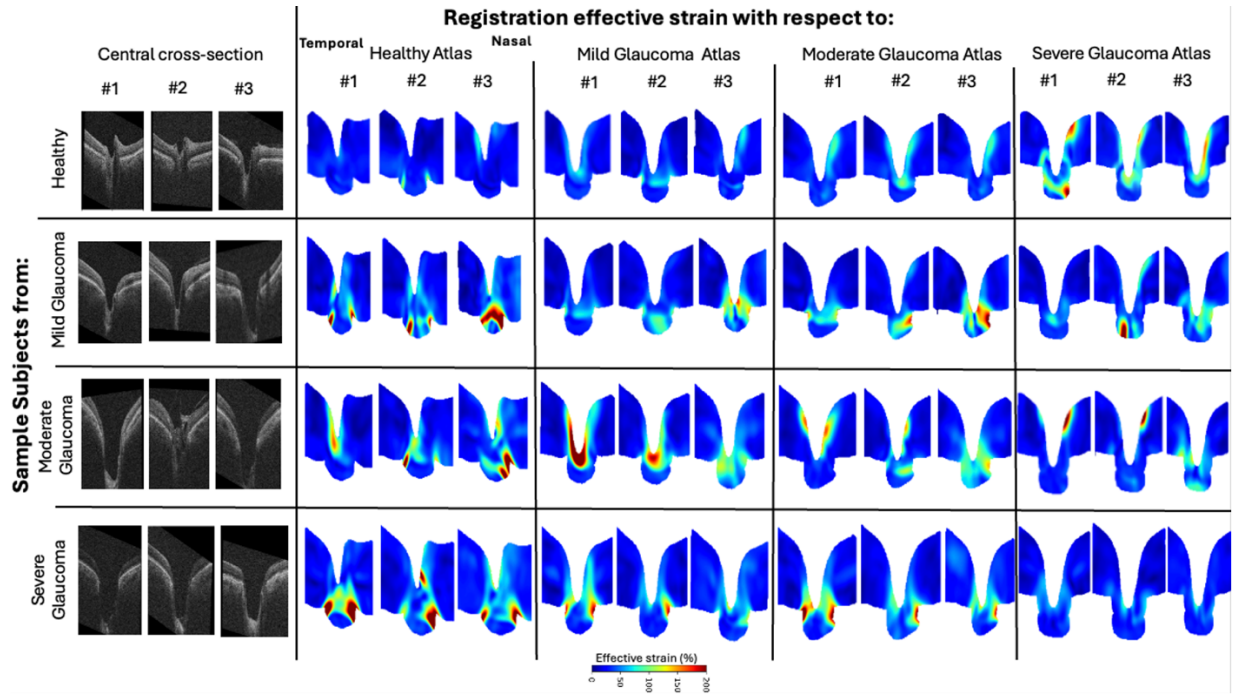

**Figure 6.** Effective strain distributions across glaucoma severity groups. Representative subjects from healthy, mild, moderate, and severe glaucoma groups (left) are shown with their registration-derived effective strain maps of a central cross-section (right), computed relative to healthy, mild, moderate, and severe atlases. Strain concentrations are consistently observed at the neuroretinal rim, lamina cribrosa insertion, and prelaminar regions, highlighting key biomechanically sensitive sites. In healthy eyes, deviations emerge primarily when registered to glaucoma atlases, whereas in severe glaucoma eyes, larger gradients of differences appear across all atlas comparisons, reflecting increasing departure from the corresponding atlas group with disease severity.